\documentclass[twoside]{ilcws10}
\usepackage[latin1]{inputenc}
\usepackage[dvips]{graphicx,epsfig,color}
\usepackage{wrapfig,rotating}
\usepackage{amssymb,amsmath,array}

\begin{document}
\title{
%%%%   Paper title goes here  %%%%%%%%%%%%%%
Measurement of Higgs Anomalous Coupling with $H \to WW^{*}$ at International Linear Collider} %% 
%***********************************************************************
% AUTHORS INFORMATION AREA
%***********************************************************************
\author{Yosuke Takubo$^1$, Katsumasa Ikematsu$^2$, Nobuchika Okada$^3$, 
Robert N. Hodgkinson$^4$,\\ and Keisuke Fujii$^{2}$
% Optional short acknowledgment: remove next line if non-needed
%\thanks{This is an optional funding source acknowledgment.}
% DO NOT MODIFY THE FOLLOWING '\vspace' ARGUMENT
\vspace{.3cm}\\
% Addresses and institutions (remove "1- " in case of a single institution)
1- Department of Physics, Tohoku University, Sendai, Japan
%% Remove the next three lines in case of a single institution
\vspace{.1cm}\\
2- High Energy Accelerator Research Organization (KEK), Tsukuba, Japan
\vspace{.1cm}\\
3- Department of Physics and Astronomy, The University of Alabama, Alabama, Japan
\vspace{.1cm}\\
4- Departamento de F\`isica Te\`orica and IFIC, Universitat de Val\`encia--CSIC, Val\`encia, Spain
}
%%***********************************************************************
% END OF AUTHORS INFORMATION AREA
%***********************************************************************

\maketitle

\begin{abstract}
The measurement of the Higgs coupling to $W$ bosons is an important program at the international linear collider (ILC) to search for the anomaly in the coupling to the gauge bosons. We study the sensitivity of ILC to the Higgs anomalous coupling to $W$ bosons by using $ZH \to \nu \nu WW^{\ast}$ events. In this article, we report the status of the study.
\end{abstract}

\section{Introduction}
The precise measurements of the Higgs boson properties are crucial to establish the theory of the electroweak symmetry breaking. Especially, the measurement of the Higgs coupling to gauge bosons is important to search for the effect of new physics. We consider the Higgs anomalous coupling to $W$ bosons in this report. The general coupling of the Higgs boson to two $W^{\pm}$ gauge bosons which is consistent with both Lorentz and gauge symmetries can be parametrised  as, 
\begin{equation}
L_{\mathrm{HWW}} = 2 M_{\mathrm{W}}^{2} \left( \frac{1}{v} + \frac{a}{\lambda} \right) H W^{+}_{\mu} W^{-\mu} + \frac{b}{\lambda} H W^{+}_{\mu\nu} W^{- \mu\nu} + \frac{\tilde{b}}{\lambda} H \epsilon^{\mu \nu \sigma \tau} W^{+}_{\mu\nu} W^{-}_{\sigma \tau},
\end{equation}
where $M_{\mathrm{W}}$ is the mass of the $W$ boson, $W^{\pm}_{\mu\nu}$ is the usual gauge field strength tensor, $\epsilon ^{\mu\nu\sigma\tau}$ is the Levi-Civita tensor, $v$ is the vacuum expectation value of the Higgs field, $a$, $b$, and $\tilde{b}$ are real dimensionless coefficients and $\lambda$ is a cutoff scale. None-zero values of $a$, $b$, and $\tilde{b}$ imply the existence of anomaly; $a$ is a scale factor to the standard model Higgs coupling, $b$ is the coefficient of a CP-even term, and $\tilde{b}$ is that of a CP-odd term. 

\begin{wrapfigure}{r}{0.45\columnwidth}
%\begin{figure}
\centerline{\includegraphics[width=0.43\columnwidth]{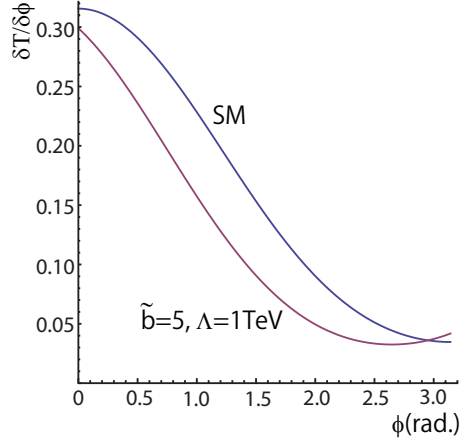}}
\vspace{-0.3cm}
\caption{The theoretical calculation of the angle between two up-type quarks coming from hadronic decays of $W$ bosons.}
\label{fig:cal}
\vspace{-0.2cm}
\end{wrapfigure}

Figure \ref{fig:cal} shows the theoretical calculation of the angle between two up-type quarks ($u$ or $c$ quark) coming from hadronic decays of $W$ bosons. If we have a finit $b$- or $\tilde{b}$-term, the cross-section and distribution shape change, whereas only the cross-section varies with the $a$-term. In this study, we focus on studying the sensitivity of ILC to the $a$ and $\tilde{b}$ by using the angular distribution of two up-type quarks.

\section{Simulation setup}
We study the sensitivity to Higgs anomalous coupling by using $ZH \to \nu \nu WW^{\ast}$, where the Higgs mass is assumed to be 120 GeV. At this Higgs mass, the branching ratio of $H \to WW^{*}$ is 15.0\%. The center of mass energy is set to be 250 GeV with an integrated luminosity of 250 fb$^{-1}$. 4-fermion final states ($\nu \nu qq$, $qq \ell \nu$, $\ell \ell \ell \ell$, $qq \ell \ell$, $qqqq$), major part of which is coming from $WW$ and $ZZ$ productions, are considered as the SM background. In order to suppress the $WW$ background, which would otherwise dominate in this study, we use 80\% right-handed polarization for the electron beam and 30\% left-handed polarization for the positron beam. 

The signal and background events were generated by WHIZARD. We used Mokka \cite{mokka} for the full simulation of the detector, in which \verb|ILD_00| is implemented as the detector model \cite{ild}. Pythia6.409 was used for hadronization. The event reconstruction was done by Marlin \cite{marlin}, where LCFIVertex package \cite{lcfi} is used for the flavor tagging of the jets.

\section{Analysis}
We selected the hadronic decay modes of $W$ bosons from Higgs decay as the signal events to fully reconstruct the Higgs mass. All the events are, therefore, reconstructed as 4-jet events. Then, the Higgs and on-shell $W$ masses are reconstructed by minimizing the $\chi^{2}$ function defined as,
\begin{equation}
\chi^{2} = 
\frac{(^{\mathrm{rec}}M_{\mathrm{H}} - M_{\mathrm{H}})^{2}}{\sigma_{H}^{2}} +
\frac{(^{\mathrm{rec}}M_{W} - M_{W})^{2}}{\sigma_{W}^{2}},
\end{equation}
where $^{\mathrm{rec}}M_{\mathrm{H}}$ is a reconstructed Higgs mass, $M_{\mathrm{H}}$ is a Higgs mass (120 GeV), $^{\mathrm{rec}}M_{W}$ is a reconstructed on-shell $W$ mass, $ M_{W}$ is a $W$ mass (80.4 GeV), and $\sigma_{H(W)}$ is the mass resolution for Higgs($W$). 

\begin{wrapfigure}{r}{0.45\columnwidth}
%\begin{figure}
\centerline{\includegraphics[width=0.43\columnwidth]{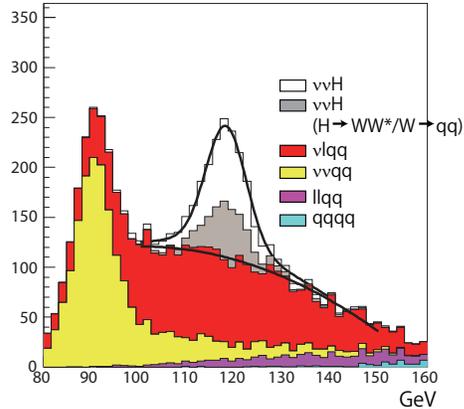}}
\vspace{-0.3cm}
\caption{Distribution of the reconstructed Higgs mass.}
\label{fig:hmass}
%\vspace{-0.2cm}
\end{wrapfigure}

After the mass reconstruction, event selection is applied to suppress background events dominating the signal events. At first, the reconstructed Higgs mass ($^{\mathrm{rec}}M_{\mathrm{H}}$) is required to be $110\ \mathrm{GeV} < ^{\mathrm{rec}}M_{\mathrm{H}} < 130\ \mathrm{GeV}$. Since a $Z$ boson decays into a neutrino pair in the signal mode, the missing mass should have a peak at the $Z$ mass. We, therefore, select events with $70\ \mathrm{GeV} < ^{\mathrm{miss}}M < 140\ \mathrm{GeV}$.

The main background in this analysis is the final states from $WW$ and $ZZ$. The angular distributions of these processes have a peak at the forward and backward region. For that reason, we require the angle of the reconstructed Higgs with respect to the beam axis ($\cos \theta_{\mathrm{H}}$) to be $|\cos \theta_{\mathrm{H}}| < 0.95$. Then, we investigate the distribution of $Y$-value, which is expected to be small to reconstruct $\nu \nu qq$ and $\nu \nu \ell \ell$ events as 4-jet events. We, therefore, select the events with $Y_{\mathrm{-}}>0.0005$, where $Y_{\mathrm{-}}$ is the threshold $Y$-value to reconstruct the events as 4 jets from 3 jets. 

After the selection cuts so far, the dominant background becomes $\ell \nu qq$. The lepton in the $\ell \nu qq$ comes from the leptonic decay of $W$ and it has larger energy than leptons from jets. We require the maximum track energy ($E_{\mathrm{trk}}$) to be below 30 GeV. 

\begin{wrapfigure}{r}{0.65\columnwidth}
%\begin{figure}
\centerline{\includegraphics[width=0.63\columnwidth]{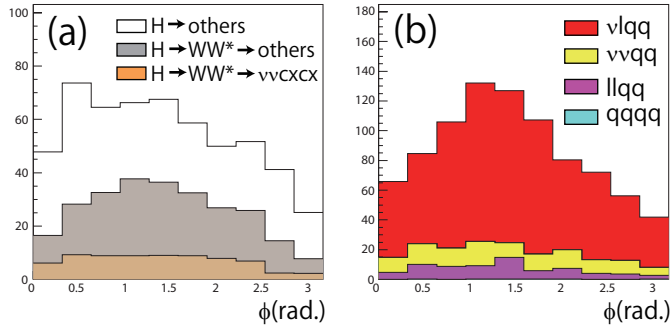}}
\vspace{-0.3cm}
\caption{Angular distribution of two c-tagged jets after the selection cuts for the signal (a) and SM background (b).}
\label{fig:plane}
\vspace{-0.2cm}
\end{wrapfigure}

Since we consider $ZH \to \nu \nu WW^{\ast}$ as the signal mode, $ZH \to \nu \nu bb$ is a background in this analysis. For that reason, $ZH \to \nu \nu bb$ events are rejected by using $b$-tagging. We require the number of $b$-tagged jets ($^{\mathrm{4-jet}}N_{b}$) to be $^{\mathrm{4-jet}}N_{b} \leq 1$. Since $ZH \to \nu \nu b$ has 2 jets in the final state, the events are reconstructed as 2-jet events for more effective $b$-tagging. Then, we select events with the number of $b$-tagged jets ($^{\mathrm{2-jet}}N_{b}$) of 0.

After all the selection cuts, we do a likelihood analysis as follows. We use $^{\mathrm{miss}}M$, $\cos \theta_{\mathrm{H}}$, $Y_{-}$, $^{\mathrm{4-jet}}N_{b}$, and the number of the charged tracks as the input variables of the likelihood function. Then, we set the likelihood cut position to maximize the signal significance. We obtain the signal significance of 7.6 with likelihood cut position of 0.79. Figure \ref{fig:hmass} shows the distribution of the reconstructed Higgs mass after all the cuts. Fitting Fig. \ref{fig:hmass} with a double Gaussian and a second order polynomial, we obtain the accuracy of BR($H \to WW^{\ast}$)  of 15.7\%, assuming that the measurement accuracy of the $ZH$ cross-section is 2.5\%. 

\begin{wrapfigure}{r}{0.45\columnwidth}
%\begin{figure}
\centerline{\includegraphics[width=0.43\columnwidth]{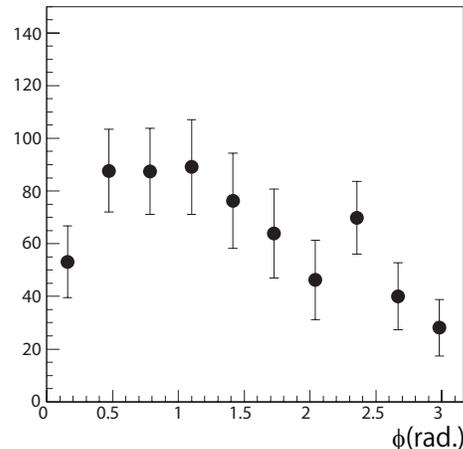}}
\vspace{-0.3cm}
\caption{ Angular distribution of two $c$-tagged jets after background subtraction.}
\label{fig:subbg}
\vspace{-0.2cm}
\end{wrapfigure}

We study to reconstruct the angular distribution of two up-type quarks from the decays of $W$ bosons. We apply double $c$-tagging to select the two up-type quarks ($c$ quark) in $ZH \to \nu \nu WW^{*} \to \nu\nu cscs$. The selection efficiency is 88\%. Figure \ref{fig:plane} shows the angular distributions of two $c$-tagged jets. Since the distribution has a peak near $\phi = 0$ rad., the distribution seems to have the angular information of two up-type quarks. The jet reconstruction, however, becomes less efficient at $\phi = 0$ and $\pi$ rad., because the separation of the jet clusters is difficult there. For that reason, the number of events at $\phi = 0$ decreases in Fig. \ref{fig:plane}. 

To obtain the distribution for $ZH$ events, we evaluate the background contamination by fitting the Higgs mass distributions for each bin of the angular distribution. After subtracting the estimated background events from Fig. \ref{fig:plane}(b), we obtain $\chi^{2}/\mathrm{ndf}$ of 0.9. We can, therefore, evaluate the background contamination within the statistical error. Subtracting the estimated background, we obtained the angular distribution of $ZH$ events as shown in Fig. \ref{fig:subbg}. For the next step, we will evaluate the sensitivity to the anomalous coupling.

\begin{table}
\centerline{\begin{tabular}{|l|r|r|r|r|r|r|r|}
\hline
& $\nu \nu H (H \to WW^{\ast} \to 4j)$ & $\nu \ell qq$ & $\nu \nu \ell \ell$ & $\ell \ell \ell \ell$ & $\nu \nu qq$ & $\ell \ell qq$ & $qqqq$ \\ \hline
No cut & 10,634 (680) & 299,866 & 103,704 & 753,964 & 63649 & 335,762 & 378,726 \\ \hline
$100 < ^{\mathrm{rec}}M_{\mathrm{H}} < 130 \mathrm{GeV}$ & 6,191 (614) & 34,540 & 6,057 & 16,561 & 2,361 & 5,488 & 518 \\ \hline
$80 < ^{\mathrm{miss}}M < 140 \mathrm{GeV}$ & 6134 (607) & 17,211 & 5,405 & 6,605 & 2,308 & 2,596 & 168 \\ \hline
$|\cos \theta_{\mathrm{H}}|<0.95$ & 5,863 (581) & 15,043 & 4,910 & 1,144 & 2,088 & 934 & 17 \\ \hline
$Y_{\mathrm{-}}> 0.0005$ 
& 5,176 (580) & 12,593 & 81 & 514 & 1,695 & 890 & 16 \\ \hline
$E_{\mathrm{trk}} < 30 \mathrm{GeV}$ & 4,826 (540) & 9,386 & 4 & 62 & 1,389 & 740 & 15 \\ \hline
$^{\mathrm{4-jets}}N_{b} \leq 1$ & 2,175 (520) & 8,692 & 4 & 46 & 1,090 & 409 & 8 \\ \hline
$^{\mathrm{2-jets}}N_{b} = 0$ & 1,518 (512) & 8,571 & 3 & 46 & 207 & 94 & 3 \\ \hline
$L > 0.79$ & 756 (348) & 1,063 & 0 & 0 & 207 & 94 & 3 \\ \hline
$N_{c} = 2$ &546 (258, cscs: 71) & 692 & 0 & 0 & 110 & 70 & 2 \\ \hline
\end{tabular}}
\caption{Cut summary.}
\label{tab:limits}
\end{table}

\section{Summary}
We have studied the sensitivity to the Higgs anomalous coupling to $W$ bosons at ILC by using $ZH \to \nu \nu WW^{\ast}$. The angular distribution of two up-type quarks from the decays of two W bosons has information of the Higgs anomalous coupling of the CP-odd term. After the selection cuts and likelihood analysis, the angular distribution can be reconstructed for $ZH$ events. For the next step, we will evaluate the sensitivity to the Higgs anomalous coupling.

\section{Acknowledgments}
The authors would like to thank all the members of the ILC physics subgroup \cite{subgroup} for useful discussions. This work is supported in part by the Creative Scientific Research Grant (No. 18GS0202) of the Japan Society for Promotion of Science and the JSPS Core University Program.

% ****************************************************************************
% BIBLIOGRAPHY AREA
% ****************************************************************************

\begin{footnotesize}
% IF YOU DO NOT USE BIBTEX, USE THE FOLLOWING SAMPLE SCHEME FOR THE REFERENCES
% ----------------------------------------------------------------------------

% ----------------------------------------------------------------------------

\end{footnotesize}

% ****************************************************************************
% END OF BIBLIOGRAPHY AREA
% ****************************************************************************

\end{document}